# Substrate-Voltage-Controlled Temporal Nonlinearity in Ferroelectric FET-based Reservoir Computing

Eishin Nako*, Ryosho Nakane, Mitsuru Takenaka, Kasidit Toprasertpong, and Shinichi Takagi

Department of Electrical Engineering and Information Systems, School of Engineering, The University of Tokyo, Bunkyo, Tokyo 113-8656, Japan

E-mail: nako@mosfet.t.u-tokyo.ac.jp

**Abstract**

**Physical reservoir computing exploits inherent nonlinearity and short-term memory of physical dynamics to achieve efficient processing of time-series data with extremely-low training cost. In this study, we demonstrate a ferroelectric field-effect transistor (FeFET)-based reservoir computing system with augmented temporal and spatial nonlinearity by utilizing both gate and substrate terminals as inputs. The ferroelectric polarization state in the next time step can additionally be controlled by modifying the electric field distribution in the gate stack of FeFET through a substrate input, enabling more diverse internal states compared with the case where inputs are applied only to the gate. To introduce a nonlinearity in the time domain, we introduce a delay between a gate input and a substrate input, which facilitates efficient nonlinear mixing between the current and past inputs. As a result, both the short-term memory and nonlinearity of the FeFET reservoir computing system are enhanced with an improved capability of feature extraction of complex input time-series. These findings demonstrate that introducing substrate input provides an additional degree of freedom for controlling ferroelectric polarization dynamics, enabling a flexible, energy-efficient, and highly integrable FeFET-based reservoir computing platform suitable for diverse time-series processing applications.**

**Introduction**

Reservoir computing (RC)[1,2] is a computational framework derived from recurrent neural networks. In RC, only the weights connected to the readout layer are adjusted by training, while the recurrent dynamics inside the reservoir remain fixed. This architecture significantly reduces the computational cost of training and has been successfully applied to tasks such as time-series prediction, pattern recognition, and nonlinear system modeling. When the reservoir is implemented using physical systems that inherently exhibit nonlinearity and short-term memory, the approach is referred to as physical RC[3,4]. By outsourcing the nonlinear computation to physical dynamics, physical RC is expected to achieve a compact RC system with high energy efficiency. On the other hand, the computational performance of physical RC is highly dependent on the diversity of the physical system's response to input signals. Accordingly, improving the nonlinearity and short-term memory of the physical dynamics is a crucial issue. Various physical platforms have been explored, including photonic systems[5], spintronic devices[6], and memristors[7]. Among them, CMOS-compatible devices are especially promising due to their scalability and integrability.

Ferroelectric field-effect transistors (FeFETs) are attractive candidates for physical RC because ferroelectric polarization dynamics and charge-polarization interactions provide short-term memory and nonlinearity. In our previous studies[8–10], FeFET-based RC was experimentally demonstrated by converting time-series data into voltage pulses applied to the gate terminal, and using the current waveforms from the drain, source, and substrate terminals as virtual nodes. This method successfully exploited dynamics in FeFETs; however, utilizing only the gate terminal as input restricts the device response to the simple gate-to-drain transfer function of FeFETs, limiting the accessible polarization dynamics and preventing full exploitation of the rich nonlinear behavior inherent to FeFETs.

It is well known that applying a voltage to the substrate terminal induces a substrate bias effect[11], leading to a modulation of electric field and carrier concentration in the channel. One of the well-known effect is a shift in threshold voltage. Thus, substrate voltage can be used to modulate the internal state of a FeFET, including its polarization dynamics, and thereby alter the drain and source currents. Motivated by this property, we employ the substrate terminal as another input terminal in addition to a gate input terminal in this study. As the substrate voltage directly controls the potential of the semiconductor, the influence of substrate voltage on the dynamics in a FeFET is essentially different from the influence of gate voltage, suggesting that introducing the substrate voltage potentially enhances the state diversity in a FeFET RC system. This work demonstrates the operation of such a dual-input FeFET RC system and provide insights into the physical origin of substrate voltage-driven dynamics in a FeFET. Moreover, the prior work[8] demonstrated that parallelizing FeFETs with delayed gate inputs improved short-term memory. In this study, we extend this approach by introducing delayed substrate input as an additional degree of freedom. Particularly, we propose a parallelization approach in which multiple FeFETs are configured with independently-delayed gate and substrate inputs to further leverage both the spatial and temporal nonlinearity of a gate/substrate dual-input FeFET RC system. This approach is expected to enable scalable and reconfigurable FeFET RC systems whose configurations can be flexibly tailored to various applications. To evaluate the impact of this approach, we employ the NARMA-10 prediction task that requires both relatively large memory capacity and strong nonlinearity.

**Experimental setting**

Figure 1 shows a cross-sectional view of an FeFET used in this work. An n-channel FeFET was fabricated on a p-type (100) Si substrate implanted with phosphorus to form the source and drain regions. The gate stack consists of Al/TiN/$Hf_{0.5}Zr_{0.5}O_2$/$SiO_x$/Si, where $SiO_x$ (0.7 nm) was formed by chemical oxidation, $Hf_{0.5}Zr_{0.5}O_2$ (10.5 nm) by atomic layer deposition, TiN (16 nm) by physical sputtering, and Al (100 nm) by thermal evaporation. The ferroelectricity of $Hf_{0.5}Zr_{0.5}O_2$ was activated by post-metallization annealing at 400°C. We used FeFETs with a channel length of 10 μm and a channel width of 100 μm. The fabricated FeFET, whose characteristics were described in Ref. 12, exhibits a memory window of 1.6 V[12].

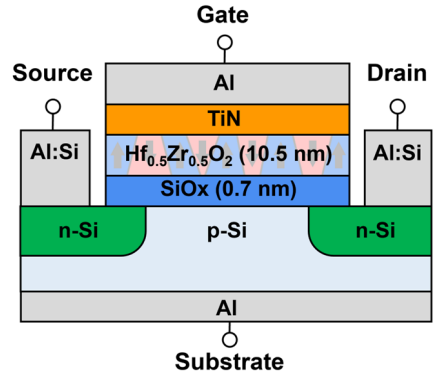

Fig.1 Cross-sectional view of FeFET.

To observe how the polarization of the FeFET changes with the substrate voltage, Fig. 2(a) shows the measured $P$–$V_g$ characteristics. The drain and source voltages were fixed at 0 V, and the substrate voltage was biased with either 0 V or –5 V. A triangular wave signal with a frequency of 125 kHz was applied to the gate electrode, and the polarization was obtained by integrating the sum of the currents flowing through the other three electrodes[13]. Compared the results of 0 V and –5 V substrate voltage, it was found that the application of a substrate voltage makes polarization switching more difficult. To gain a better understanding of the reduced polarization switching, we examined the impact of substrate voltage on the switching polarization $P_{sw}$ in each gate voltage polarity by using the measurement setting in Fig. 2(b). After fully polarizing downward by applying a positive gate voltage, a negative triangular pulse with a width of 4 μs was applied to the gate under a fixed substrate voltage, and the sum of the currents through the other electrodes was integrated to obtain the negative switching polarization. The positive switching polarization was measured similarly by reversing the polarity of the applied gate voltage. Figure 2(c) shows how the positive and negative switching polarizations are affected by the substrate voltage. The positive switching polarization remains almost constant regardless of the substrate voltage, whereas the negative switching polarization decreases significantly when the substrate voltage drops below –1 V.

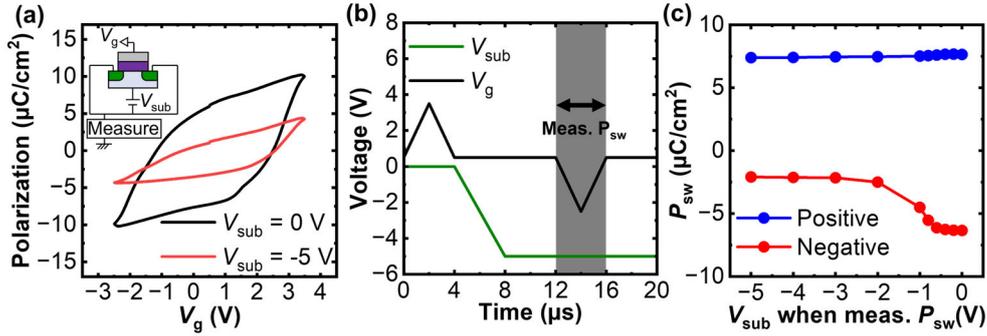

Fig.2 (a) Schematic diagram of the $P$–$V_g$ measurement and changes in $P$–$V_g$ characteristics with substrate voltage. (b) Measurement sequence of the switching polarization $P_{sw}$ for negative gate pulse. (c) Variation of the switching polarization $P_{sw}$ with substrate voltage.

The reason for this asymmetry is explained in the following based on the MOS band structure of the FeFET. Figure 3 illustrates the MOS band diagrams with metal/ferroelectric/interfacial layer/Si structure under positive/negative gate voltages and zero/negative substrate voltages. Figure 3(a) shows the band diagram of a positive gate voltage and 0 V substrate voltage. When the gate voltage is higher than the threshold voltage needed to form an electron inversion layer, the Si surface is in inversion. In this condition, the gate overdrive voltage is mostly applied across the ferroelectric/interfacial layer structure, resulting in a large electric field across the ferroelectric layer. For this reason, the electric field across the ferroelectric layer can easily exceed the coercive field and induce polarization switching. Figure 3(b) shows the diagram with a negative substrate voltage. Although the depletion layer widens due to the substrate bias effect and the threshold voltage slightly increases, the transistor remains the on-state for an applied gate voltage sufficiently larger than the threshold voltage. Thus, the electric field across the ferroelectric layer is also still sufficient to switch the ferroelectric polarization. In other words, the impact of substrate voltage on the behavior of positive polarization switching is relatively small. Next, we consider of the operation under negative gate voltage. Figure 3(c) shows the MOS band diagram at a negative gate voltage and 0 V substrate voltage. In this condition, the Si surface is in accumulation, so most of the applied gate voltage drops across the ferroelectric/interfacial layer structure. Then, as similar to the inversion condition, the electric field in the ferroelectric layer can easily exceed the coercive field of the ferroelectric layer and leads to polarization switching.

On the other hand, the behavior is completely different for the condition of negative gate and negative substrate voltages, as shown in Fig. 3(d). Applying a negative substrate voltage raises the energy level of the Si bands in the neutral region along with the quasi-Fermi level for holes ($E_{Fp}$). When the negative substrate voltage is sufficiently large, $E_{Fp}$ moves far away from the Si valence band, and as a result, the hole accumulation layer at the surface is no longer formed with a depletion layer formed in the Si layer. This disappearance of holes near the interface causes the electric field to extend into the Si substrate, thereby essentially reducing the field across the ferroelectric/interfacial layer. Consequently, the ferroelectric polarization becomes less likely to switch at this low electric field. Hence, only in this condition the polarization switching can be effectively suppressed by substrate voltage, agreeing with the results observed in Fig. 2(b). These results indicate that the polarization switching dynamics can be partially controlled by the substrate voltage. Furthermore, this control by the substrate voltage is not a simple linear superposition of the control by the gate voltage: the electric field across ferroelectric is a complex function of the gate and substrate voltages. When applied to FeFET reservoir computing, using substrate input alongside gate input enables partial control of polarization dynamics, enhancing the diversity of polarization states as well as the degrees of freedom of the reservoir internal states, which can improve nonlinearity and short-term memory capacities.

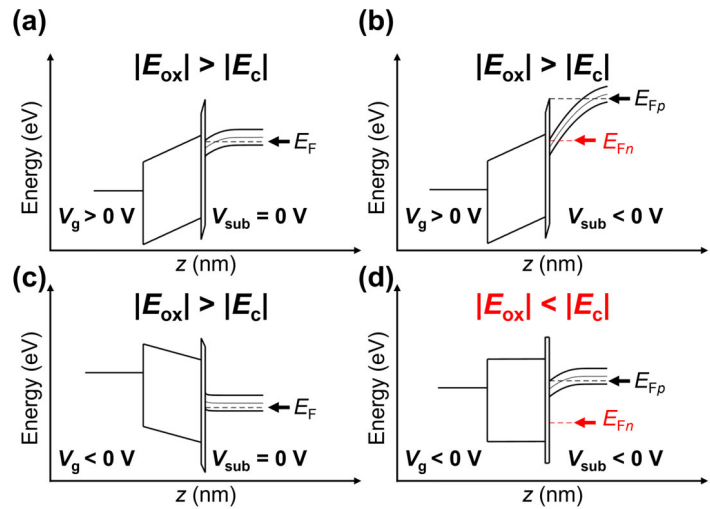

Fig.3 Band diagrams of MOS structure under applied substrate bias: (a) positive gate voltage with 0 V substrate voltage, (b) positive gate voltage with negative substrate voltage, (c) negative gate voltage with 0 V substrate voltage, and (d) negative gate voltage with negative substrate voltage. Only in case (d) does the polarization undergo partial switching. Note that the Si region is scaled down by a factor of 1/100 to improve the visibility of the band bending in Si.

**FeFET reservoir computing system with Vsub input**

In our previous studies[8–10], the FeFET RC system received inputs only from the gate terminal. In this work, however, inputs are also applied to the substrate terminal. A schematic diagram of the constructed FeFET RC system is shown in Fig. 4. The digital time-series input data are converted into a sequence of triangular pulses such that one time step of the input corresponds to a single triangular pulse with a width of 4 μs.

When applying input through the gate terminal, the center voltage is set to 0.5 V and the pulse amplitude to ±3 V, which is confirmed to be sufficient to induce ferroelectric polarization switching for zero substrate voltage. For the substrate terminal, the center voltage is set to –2.5 V with a pulse amplitude of ±2.5 V. For gate input, an input value of "0" is converted into a negative pulse and "1" into a positive pulse. For substrate input, an input value of "0" is converted into a positive pulse toward zero substrate voltage and "1" into a negative pulse toward larger negative substrate voltage. The drain voltage of the FeFET was fixed at 0.3 V, while the source voltage was fixed at 0 V. The drain and source current waveforms show complex time evolution due to the dynamics of ferroelectric polarization driven by the gate and substrate voltages. These currents were sampled at an interval of 40 ns. Since the time step of digital input is 4 μs, this results in 100 measurement points per terminal for each time step, and thus 200 measurement points in total when combining both drain and source currents. These measurement points were concatenated and used as virtual nodes[14], forming an internal state vector $\mathbf{x}(n)$ with $N = 200$ virtual nodes at $n$-th time step. We also examined the conditions that voltage pulses applied to the substrate terminal were delayed by $\tau_{\text{sub}}$ time steps relative to those applied to the gate terminal.

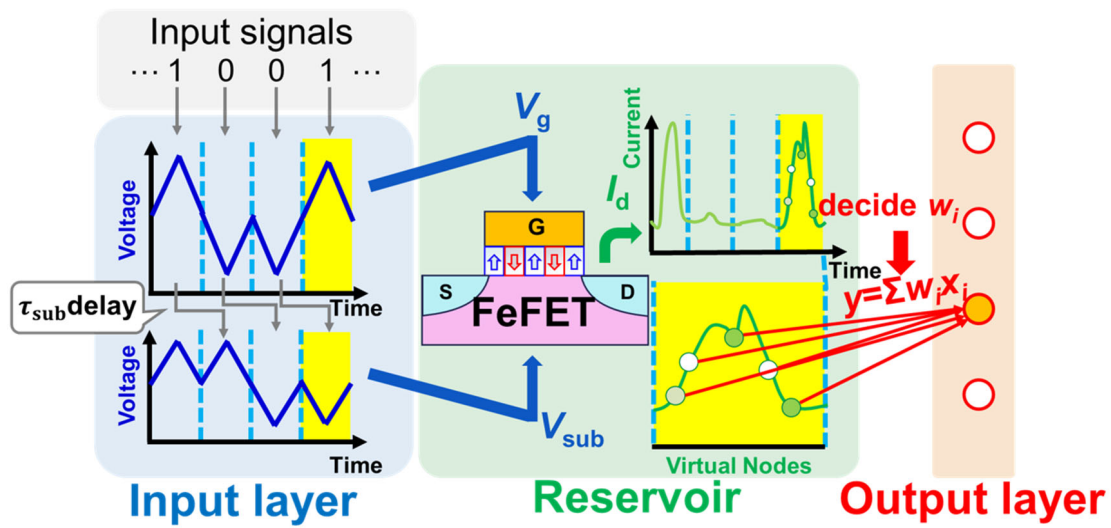

Fig.4 Schematic diagram of the FeFET RC system with inputs applied from both the gate and substrate terminals.

In this experiment, ten different sequences of 1000-step time-series data were generated and input to the FeFET. The first 500 time steps of each sequence were washed-out, and the subsequent 500 time steps were employed for training and inference. Five subsets were generated by pairing two sequences, with four subsets ($L$ = 4000 time steps) used for training and one subset ($L'$ = 1000 time steps) for inference.

During training, under an input sequence $u(n)$ of $L$ time steps, ridge regression[15] was performed using the $N \times L$ matrix **X**, constructed from the internal state vector **x**($n$) over the $L$ time steps. Here, $\lambda$ denotes the ridge parameter, $\hat{\mathbf{y}}$ the target vector of size $L$ for the RC system, **w** the weight vector of size $N$, and **I** the identity matrix. The value of $\lambda$ was chosen based on performance evaluation using the training data, selecting the value that achieved the highest performance.

$$\mathbf{w} = \hat{\mathbf{y}} \mathbf{X}^\mathrm{T} (\mathbf{X}\mathbf{X}^\mathrm{T} + \lambda \mathbf{I})^{-1}.$$

During inference, the unseen input sequence $u'(n)$ of $L'$ time steps was fed to the trained FeFET RC system. The output vector **y** of the RC system was obtained by calculating the inner product of the $N \times L'$ matrix **X′**, generated from the internal state vector over $L'$ time steps, with the weight vector **w** obtained during training.

$$\mathbf{y} = \mathbf{w}\mathbf{X}'.$$

The performance was evaluated using the short-term memory (STM) task[16][17] (also known as a delay task) and the temporal exclusive OR (XOR) task[17], which are the standard benchmarking tasks. The STM task estimates the input $d$ time steps earlier, thereby assessing short-term memory capacity. The temporal XOR task computes the XOR of the present input and the input at $d$ time steps earlier, thereby evaluating both short-term memory and nonlinearity. The target vectors for the STM and XOR tasks are expressed as:

$$\hat{y}_\mathrm{MC}(d) = u(n-d),$$
$$\hat{y}_\mathrm{XOR}(d) = \mathrm{XOR}\big(u(n), u(n-d)\big).$$

The coefficient of determination $r^2$ between the target output $\hat{\mathbf{y}}(d)$ and the RC output $\mathbf{y}(d)$ is calculated, and the computational capacity $C_\mathrm{MC}$ or $C_\mathrm{XOR}$ is given by $\sum_{d=1}^{D} r^2(d)$. Here, $D$ was chosen sufficiently large such that $r^2(d)$ became zero, and $D$ was set to 9 time steps. The performance of the RC system was evaluated using this computational capacity. Figure 5 shows the computing capacities $C_\mathrm{MC}$ and $C_\mathrm{XOR}$ for the conditions without substrate input and with substrate input delayed by $\tau_\mathrm{sub}$ time steps. Compared to the conditions without substrate input, no significant difference is observed when pulses are applied to the substrate without delay, implying that feeding the same input to the gate and substrate terminals does not contribute to the improvement of short-term memory and nonlinearity of the FeFET reservoir. However, as the delay $\tau_\mathrm{sub}$ increases from 1 to 3 time steps, the capacities of both tasks improve, while further increases in $\tau_\mathrm{sub}$ result in slight capacity drop. These results demonstrate that applying proper delayed inputs to the substrate terminal leads to temporal nonlinearity and thus enhances both the short-term memory and nonlinearity required for performing RC.

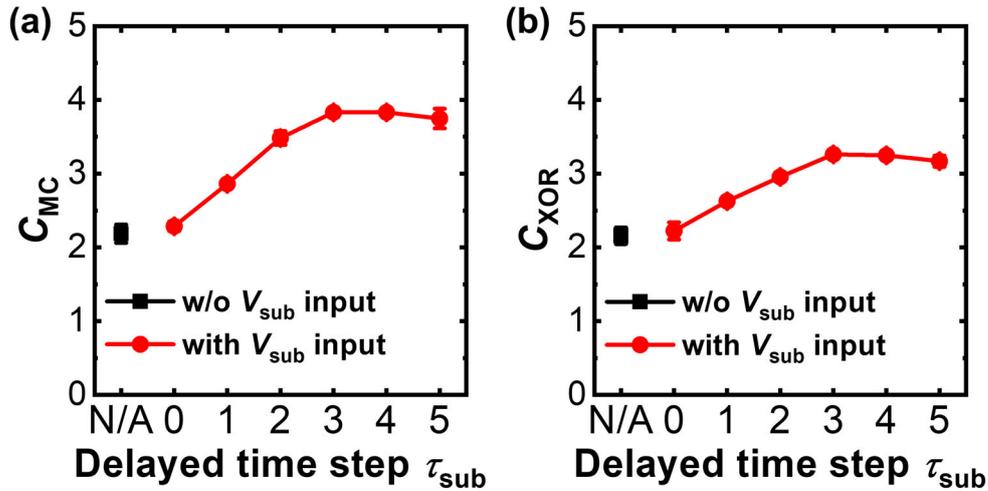

Fig.5 Performance evaluation of the FeFET RC system with delayed substrate input: (a) short-term memory task and (b) temporal XOR task.

To clarify the reason why delayed substrate input contributes to the improved system performance, the current waveforms of the FeFET were analyzed. Figure 6 compares the drain current waveforms when the input sequences of the last three steps ($u(n-2)$, $u(n-1)$, $u(n)$) are (0,0,0) and (1,0,0). Here, we compare the current waveforms when the present inputs are the same but the past inputs are different. As the drain current waveform is the main component forming the internal state vector $\mathbf{x}(n)$, the difference between these two waveforms should reflect how strongly the information from two time steps earlier ($n$-2) influence the internal state vector $\mathbf{x}(n)$ at the present step. Figure 6(a) shows the drain current waveform without substrate input, while Fig. 6(b) shows that with substrate input delayed by one time step. The output current waveforms without substrate input for (0,0,0) and (1,0,0) are nearly identical, making it difficult to distinguish the past input. In contrast, the waveforms with delayed substrate input differ significantly depending on the input history.

When there is no substrate input, the FeFET exhibited the behavior shown in Fig. 6(c) for the input sequence (1,0,0). When "1" is applied at step $n$-2, a positive voltage pulse is applied to the gate terminal and set the polarization to be oriented downward. Then, $u(n-1) =$ "0" or a negative voltage pulse switched the polarization to the upward state. By the time the final "0" is applied, the polarization is already oriented upward, so polarization dynamics are expected to be minimal. For the sequence (0,0,0), the polarization is also kept at the upward state, resulting in no switching events. For this reason, the waveforms observed at the present step $n$ is the non-switching current waveforms of FeFET for both (0,0,0) and (1,0,0), resulting in the similar waveforms with only little difference.

On the other hand, the behavior is different when a one-step delayed substrate input is applied, as shown in Fig. 6(d). As observed in Figs. 3(a) and (b), a positive gate voltage pulse orients the polarization downward at the step $n$-2, regardless of the substrate voltage. Next, at the step $n$-1, $u(n$-

1) = "0" or a negative voltage pulse is applied to the gate terminal while delayed input $u(n-2)$ = "1" or a -5 V pulse is applied to the substrate terminal. As explained in Fig. 3(d), the Si layer tends to remain at the depletion condition even at a negative gate voltage if the negative substrate voltage is biased. Consequently, the electric field across the ferroelectric layer is insufficient for complete switching, resulting in partial polarization switching after the step $n$-1. At this condition, a significant portion of the polarization still remains at the downward state. At the final step $n$, both the $u(n)$ = $u(n-1)$ = "0" corresponds to a negative voltage pulse to the gate terminal while the substrate voltage is nearly zero. In this way, the polarization switches upward, generating polarization switching current responsible for the peak observed in Fig. 6(b). For the (0,0,0) sequence, the polarization remains at the upward state and no polarization switching occur. As a result, the output current waveforms for the (1,0,0) and (0,0,0) sequences become distinct under delayed substrate input, showing the polarization switching behavior and non-switching behavior, respectively.

As a result, employing delayed input to the substrate terminal enhances the difference in the current waveforms under different input sequences, allowing the input history to be effectively distinguished. These observations indicate that applying an appropriate substrate input alongside the gate input can partially suppress gate-induced polarization switching depending on the input sequence, allowing the FeFET RC system to exhibit more diverse responses to different inputs.

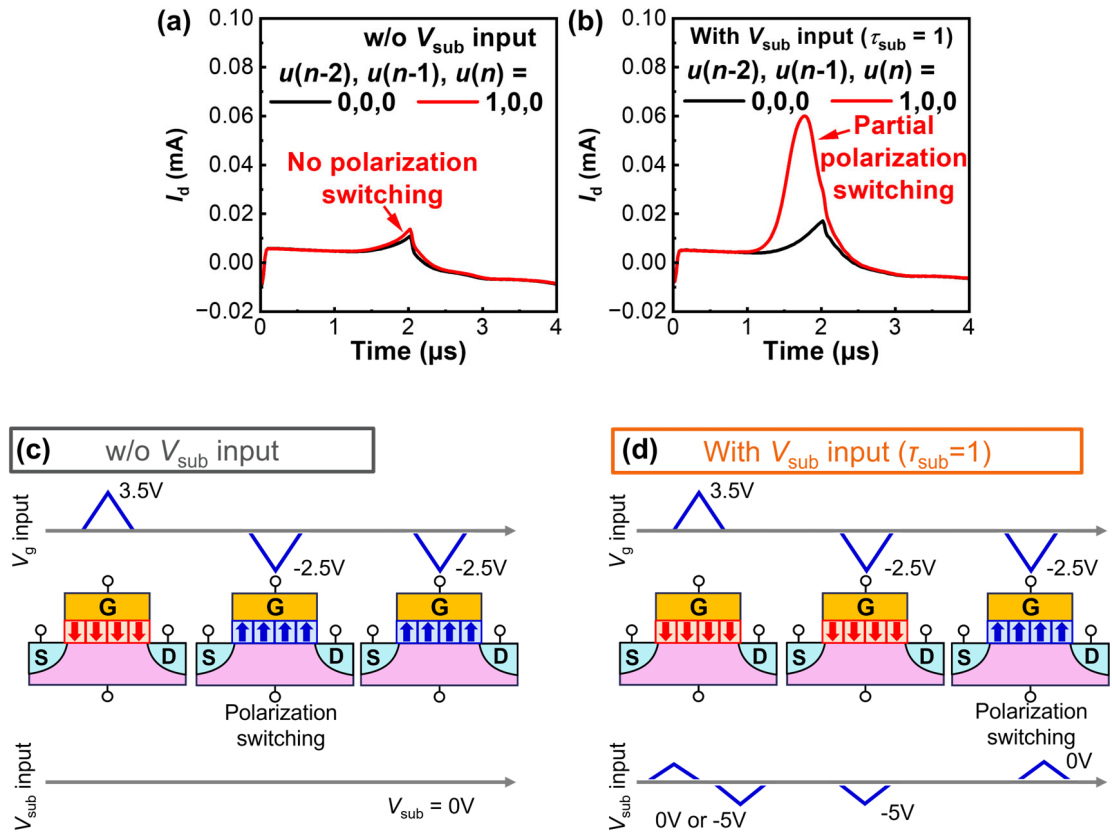

Fig. 6 Comparison of drain current waveforms for input sequences of the last three steps are (0,0,0) and (1,0,0): (a) without substrate input and (b) with substrate input delayed by one time step. Polarization behavior of FeFET when input sequence is (1,0,0) (c) without $V_{sub}$ input or (d) with 1 time step delayed $V_{sub}$ input.

**Parallel FeFET reservoir computing system with $V_{sub}$ input**

In our previous study, we explored the construction of a larger FeFET RC system by parallelizing multiple FeFET subsystems with different input delays (gate delay method)[8], as illustrated in Fig. 7(a). However, the nonlinearity is not expected to be improved in the gate delay method, as this parallelizing scheme does not introduce any new nonlinear interaction to the system. To overcome this limitation, we propose a new method named sub delay method, shown in Fig. 7(b).

- Gate delay method

Multiple FeFETs receive only gate inputs with each delayed by 0, 1, … or $\tau_{max}$ time step, while no substrate input is applied. This approach improved the short-term memory characteristics of the system as the past input is explicitly fed to the RC system.

- Sub delay method

Multiple FeFET subsystems are parallelized without any delay in the gate input, while the substrate input is delayed by 0, 1, … or $\tau_{max}$ time step for each device.

For each method, the outputs from all FeFET subsystems, with 200 virtual nodes each, are concatenated to form an internal state vector. As demonstrated in Fig. 5(b), combining the gate and substrate inputs with a delay enhances the nonlinearity; therefore, the sub delay method is expected to improve both short-term memory and nonlinearity simultaneously.

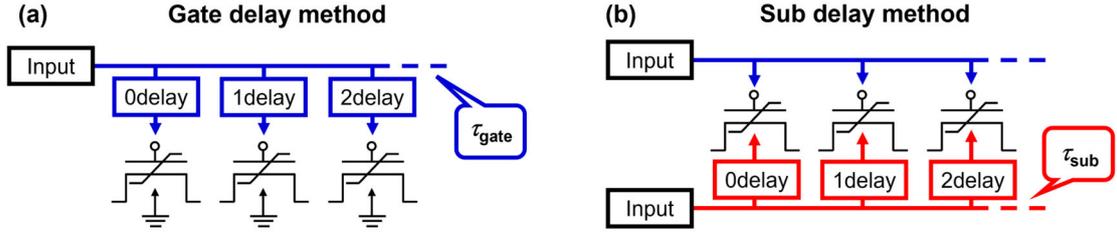

Fig.7 Two schemes for combining input conditions: (a) gate delay method: gate input delayed by 0, 1, ... time step without substrate input, (b) sub delay method: substrate input delayed and gate input without delay.

The results of the $C_{MC}$ and $C_{XOR}$ evaluations for this configuration are shown in Fig. 8. The horizontal axis represents the maximum delay time step, $\tau_{max}$. In this configuration with $\tau_{max}$, the FeFET RC system consists of $\tau_{max} + 1$ of FeFET subsystems in parallel, yielding $2 \times 100 \times (\tau_{max} + 1)$ (source/drain terminals ×sampling points per time step × parallelized devices) virtual nodes in total. As shown in Fig. 8(a), both the gate delay method and the sub delay method exhibit an increase in $C_{MC}$ as $\tau_{max}$ increases. Since the gate input has a larger influence on the polarization state and hence more directly reflects information of the input signal, the gate delay method achieves slightly higher $C_{MC}$ values. Figure 8(b) shows the evaluation results for the $C_{XOR}$ task. While the gate delay method shows little improvement in $C_{XOR}$ with increasing $\tau_{max}$ because there is no nonlinear operation in the gate delay method, the sub delay method exhibits a clear improvement in $C_{XOR}$ with an increase in $\tau_{max}$. This result suggests that the combination of gate and delayed substrate inputs enables the system output to reflect nonlinear interaction between current and past inputs. These findings indicate that parallelizing FeFETs with delayed substrate inputs enhances not only short-term memory characteristics but also nonlinearity, which is particularly beneficial for the RC system.

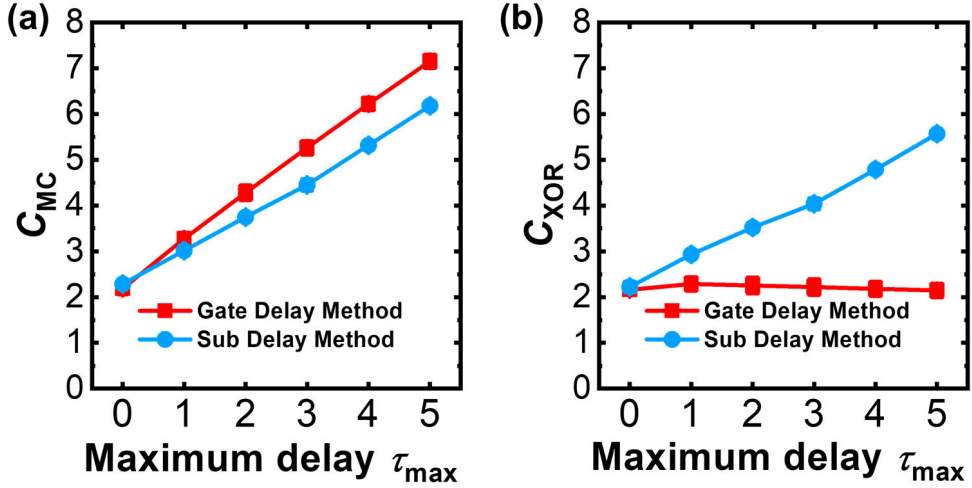

Fig.8 Performance of the FeFET RC system with Gate Delay Method and Sub Delay Method schemes: (a) short-term memory task and (b) temporal XOR task.

**Analog Parallel FeFET reservoir computing system with $V_{sub}$ input**

Through the preceding experiments, it was confirmed that substrate input can enhance both the short-term memory and nonlinearity of the RC system in basic digital tasks. In this section, we evaluate the effect of substrate input on the performance of more practical analog tasks. We suppose in this study that the time-series data consist of random analog values between 0 and 0.5. These analog values are converted into triangular pulses with a width of 4 μs. Our previous report has revealed that applying the split analog method to convert the analog input into a gate voltage pulse achieves higher computational capability.[18] Using this split analog method, the gate input is converted into triangular pulses with a center voltage of 0.5 V and a maximum pulse amplitude of 3 V with a gap of 2 V. That is, the input value from 0 to 0.25 is mapped to the pulse voltage amplitude of -3 V to -2 V while the input value from 0.25 to 0.5 is mapped to the pulse voltage amplitude of 2 V to 3 V, leaving no voltage amplitude from -2 V to 2 V to be mapped from any input values. This is because the middle voltage range is lower than the coercive voltage of the polarization and thus cannot drive polarization dynamics in FeFETs.[18] For the substrate input, the linear analog method is applied, in which the input value is linearly mapped to the pulse voltage amplitude. The center voltage is set to –2.5 V, and the pulse amplitude is varied from 2.5 V to -2.5 V corresponding to the input value from 0 to 0.5. Furthermore, we also combine two FeFETs, where one receiving the original voltage waveform and the other receiving its inverted waveform. This is intended to enhance the computational capability since FeFET operation is asymmetric with respect to a gate voltage, where the drain current exhibits poorer dynamics at a negative gate voltage due to the channel being turned off.[19] The drain and source currents were measured at an interval of 400 ns, yielding 10 virtual nodes per terminal per time step. The number of virtual nodes is decreased from that described in the last section as the number of subsystems is larger. Since each of the two FeFETs has both drain and source terminals, 40 virtual nodes are obtained for each delay condition. To enhance the design flexibility of the FeFET RC system, we further consider additional schemes in which both gate and substrate inputs are delayed, which can utilize various combinations of subsystems with delay condition ($\tau_{gate}$, $\tau_{sub}$), as shown in Fig.9.

In the experiment with analog data, 30 types of 1000-step time-series data were generated and input into the FeFETs. The first 500 time steps of each sequence were discarded for washout, and the latter 500 time steps were used for training and inference. Six sequences were combined to form one subset, resulting in five subsets in total. Among these subsets, four subsets ($L$ = 12,000 time steps) were used for training, and one subset ($L'$ = 3000 steps) was used for inference.

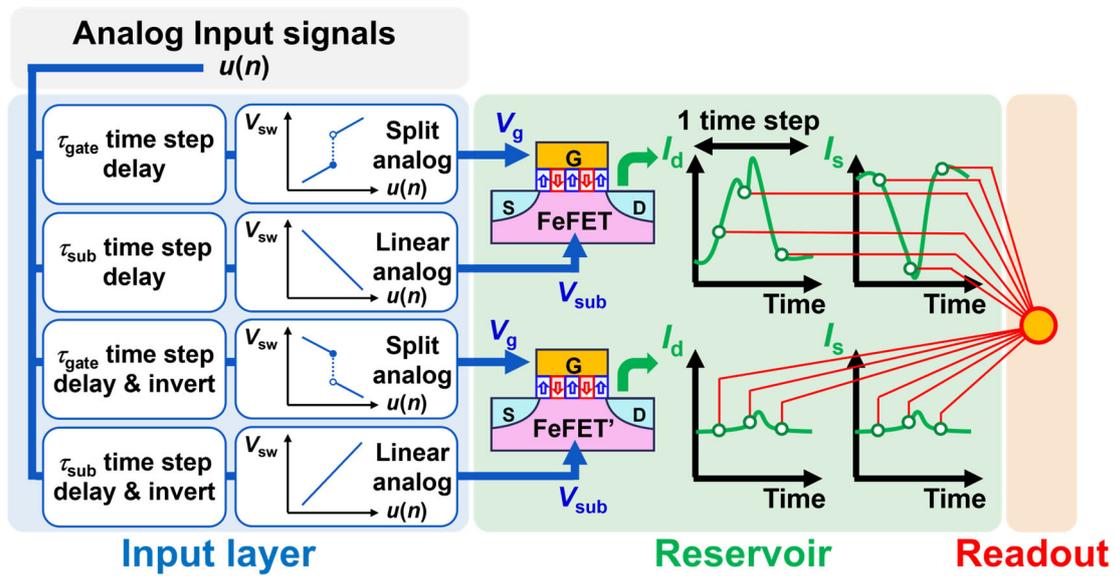

Fig.9 Schematic diagram of the FeFET RC subsystem in which an inverted input signal is applied to an additional FeFET. The gate input is delayed by $\tau_{gate}$ time steps and the substrate input by $\tau_{sub}$ time steps.

Whereas there are numerous possible ways to combine FeFET RC subsystems with delay conditions ($\tau_{gate}$, $\tau_{sub}$), this study initially evaluated the four schemes shown in Fig. 10.

- "Gate input (GI)-$\tau_{gate}$" scheme

It corresponds to the gate delay method.

- "Gate & substrate input (GSI)-$\tau_{sub}$" scheme

It corresponds to the sub delay method.

- "GSI–$\tau_{gate}$" scheme

Substrate input is applied without delay, while the gate input is delayed by increments of one time step up to $\tau_{max}$.

- "GSI–$\tau_{gate}$ & $\tau_{sub}$" scheme

Both $\tau_{gate}$ and $\tau_{sub}$ are varied from 0 to $\tau_{max}$, and all possible combinations are applied.

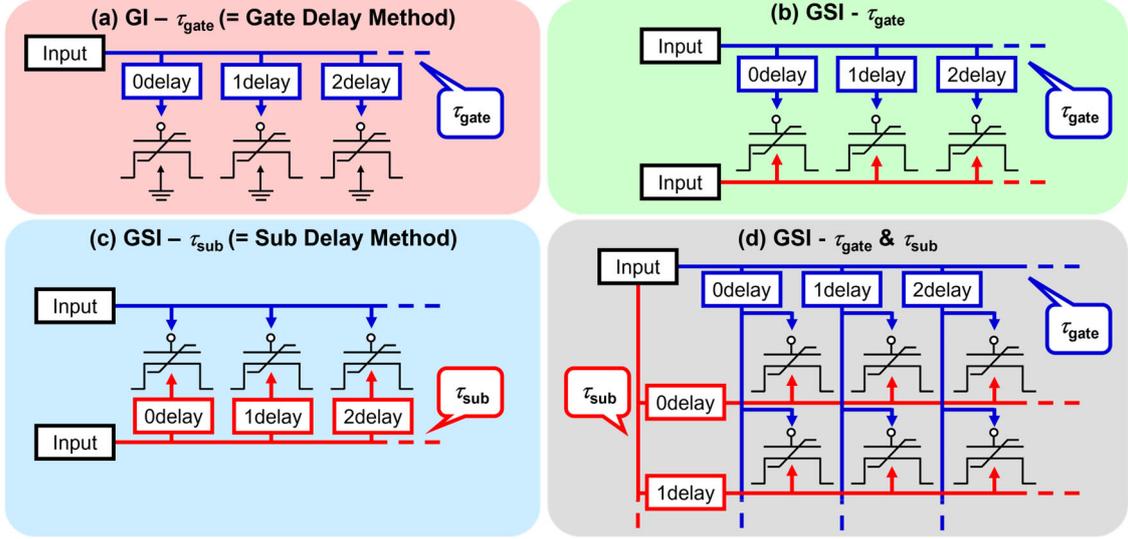

Fig.10 Four schemes for combining input conditions: (a) GI-$\tau_{gate}$: gate input delayed without substrate input, (b) GSI–$\tau_{gate}$: gate input delayed with undelayed substrate input, (c) GSI–$\tau_{sub}$: substrate input delayed with undelayed gate input, and (b) GSI–$\tau_{gate}$ & $\tau_{sub}$: all combinations of $\tau_{gate}$ and $\tau_{sub}$ from 0 to $\tau_{max}$.

The system performance was evaluated using the NARMA-$p$ prediction task[20,21]. This task requires predicting the next output of a nonlinear autoregressive moving average system of order $p$, where the task difficulty increases with larger $p$. For NARMA-2, the system is defined as:

$$\hat{y}(n) = 0.4\hat{y}(n-1) + 0.4\hat{y}(n-1)\hat{y}(n-2) + 0.6u(n)^3 + 0.1.$$

For NARMA-$p$ with $p \geq 3$, the system is defined as:

$$\hat{y}(n) = 0.3\hat{y}(n-1) + 0.05\hat{y}(n-1)\sum_{k=0}^{p-1}\hat{y}(n-k-1) + 1.5u(n-p+1)u(n) + 0.1.$$

The performance on NARMA prediction tasks is quantified using the normalized mean square error (NMSE):

$$NMSE = \frac{\frac{1}{M}\sum_{n=1}^{M}(y(n)-\hat{y}(n))^2}{\frac{1}{M}\sum_{n=1}^{M}(\hat{y}(n)-\overline{\hat{y}(n)})^2}.$$

It should be noted for NARMA-2 that some studies define the denominator as $\hat{y}(n)^2$ instead of $(\hat{y}(n)-\overline{\hat{y}(n)})^2$, and thus caution must be taken when comparing results across the different definitions.

Figure 11 compares the NMSE among the different schemes. Remind that a smaller NMSE indicates that the RC system can predict time-series data more accurately, which is the opposite to the computational capacity. In Fig. 11(a), the performance on the NARMA-10 prediction task is shown for different $\tau_{max}$. For $\tau_{max} > 10$, the NMSE of "GI-$\tau_{gate}$" does not further decrease, whereas both "GSI-$\tau_{gate}$" and "GSI-$\tau_{sub}$" achieve lower NMSE values than "GI-$\tau_{gate}$" and exhibit nearly identical performance. This improvement arises because in "GSI-$\tau_{gate}$" and "GSI-$\tau_{sub}$", polarization control is performed using the delay between the gate and substrate inputs and thus the enhancement in temporal nonlinearity is expected. The slightly better performance of "GSI-$\tau_{gate}$" is attributed to the fact that polarization control via the gate has a more direct effect on the polarization state, which allows delayed gate input to improve short-term memory more efficiently. One observation is that for a given $\tau_{max}$, "GSI–$\tau_{gate}$ & $\tau_{sub}$" always provides the lowest NMSE as it consists of more variety of subsystems. However, this method suffers from a drastic increase in the number of virtual nodes as it employs a matrix of $\tau_{gate}$ and $\tau_{sub}$. In particular, for the other three methods, the number of nodes is $4 \times 10 \times (\tau_{max} + 1)$, while for "GSI–$\tau_{gate}$ & $\tau_{sub}$", it is $4 \times 10 \times (\tau_{max} + 1)^2$. Figure 11(b) compares performance as a function of the number of virtual nodes. When normalized by node count, "GSI–$\tau_{gate}$" and "GSI–$\tau_{sub}$" exhibited lower NMSE, outperforming the other methods. Considering computational cost and memory usage, achieving lower NMSE with fewer nodes in "GSI–$\tau_{gate}$ and GSI–$\tau_{sub}$" is advantageous in actual systems.

Figure 11(c) shows the NMSE for "GSI–$\tau_{sub}$" for NARMA-2, NARMA-3, NARMA-7, and NARMA-10 prediction tasks. In all cases, sufficiently large $\tau_{max}$ leads to reduced NMSE, indicating that "GSI–$\tau_{sub}$" is effective not only for NARMA-10 but also for other NARMA prediction tasks. The NMSE improvement tends to saturate when $\tau_{max}$ approximately goes beyond the order $p$ as expected that NARMA-$p$ involves the information until $p$-1 steps earlier.

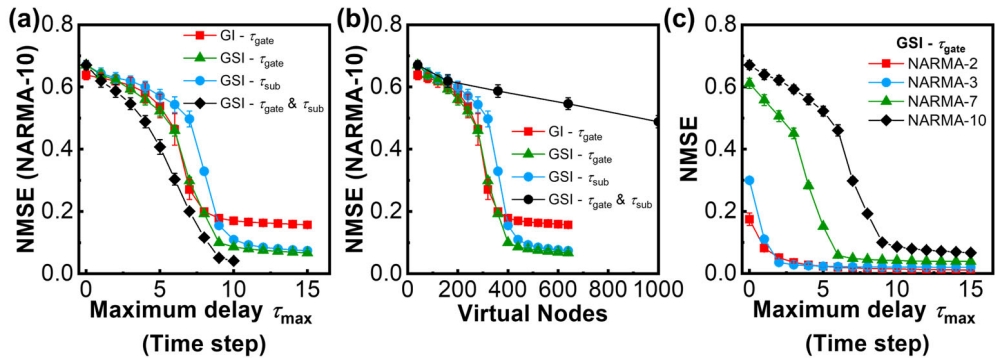

Fig.11 NMSE among the four input schemes in NARMA-10 prediction tasks: (a) NMSE as a function of maximum delay time $\tau_{max}$, where "GSI–$\tau_{gate}$ & $\tau_{sub}$" shows the smallest NMSE, (b) NMSE as a function of the number of virtual nodes, where "GSI–$\tau_{gate}$" and "GSI–$\tau_{sub}$" yield lower NMSE than other schemes at the same virtual node size, and (c) NMSE of NARMA-2, -3, -7, and -10 prediction tasks using "GSI–$\tau_{sub}$", showing that larger $\tau_{max}$ reduces NMSE for all tasks.

**Optimization for specific task**

The performance of different subsystem combination schemes has been discussed in Fig. 11. Here, we analyze how each individual delay condition shown in Fig. 9 contributes to the overall performance, and explore the possibility of systematically constructing a system with better performance. Figure 12 compares the NMSE obtained when the NARMA-10 prediction task was performed using individual FeFET RC subsystems with specific delay conditions ($\tau_{gate}$, $\tau_{sub}$). The results indicate that the performance strongly depends on the choice of the delay condition. In particular, the FeFET RC subsystems with the delay conditions ($\tau_{gate}$, $\tau_{sub}$) = (0, 9) and (9, 0) yield especially lower NMSE. These input conditions are well suited for enhancing the nonlinear interaction between the current input and the input from nine steps earlier. This interaction corresponds to the third term in the NARMA-10 equation, $1.5u(n-9)u(n)$. This fact can explain the result that combining these two FeFET RC subsystems is effective to solve NARMA-10 task.

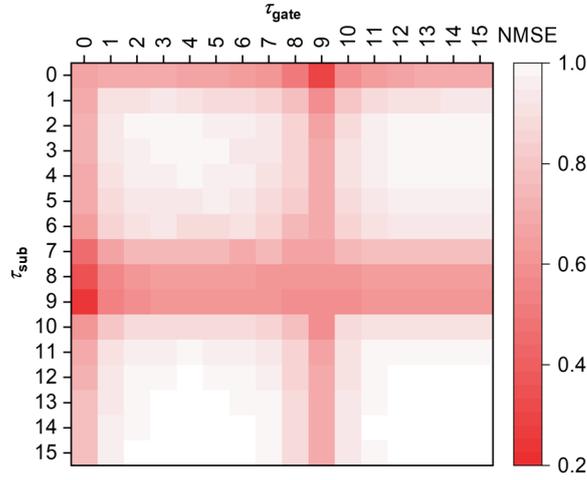

Fig.12 NMSE in the NARMA-10 prediction task for individual FeFET RC subsystems with different delay conditions.

To evaluate whether the performance can be further improved, we tested additional configuration which is denoted as "GSI–$\tau_{gate}$ & ($\tau_{gate}$, $\tau_{sub}$) = (0, 9)" illustrated in Fig. 13(a).

- "GSI–$\tau_{gate}$ & ($\tau_{gate}$, $\tau_{sub}$) = (0, 9)" scheme

The input condition $(\tau_{gate}, \tau_{sub}) = (0,9)$ is added to the "GSI-$\tau_{gate}$" scheme, which achieved a low NMSE value with a relatively small number of virtual nodes.

For a fair discussion, we made a comparison between systems with the same total number of subsystems. Specifically, for a given $m$, the comparison was made between the conventional "GSI-$\tau_{gate}$" scheme with $\tau_{max} = m$ and the "GSI-$\tau_{gate}$" scheme with $\tau_{max} = m - 1$ in parallel with the $(\tau_{gate}, \tau_{sub}) = (0,9)$ configuration. In both configurations, the total number of delay combinations is $M = m + 1$.

As shown in Fig. 13(b), the inclusion of $(\tau_{gate}, \tau_{sub}) = (0, 9)$ enables the system to achieve lower NMSE with the same number of delay conditions, thanks to the contribution from the $(\tau_{gate}, \tau_{sub}) = (0, 9)$ configuration. At $M = 10$, the two methods exhibit nearly identical performance. This is because $(\tau_{gate}, \tau_{sub}) = (9, 0)$ is included in the "GSI–$\tau_{gate}$" scheme, whereas $(\tau_{gate}, \tau_{sub}) = (0, 9)$ is included in the "GSI–$\tau_{gate}$ & $(\tau_{gate}, \tau_{sub}) = (0, 9)$" scheme. For $M \geq 11$, both $(\tau_{gate}, \tau_{sub}) = (9, 0)$ and $(0, 9)$ are included in the "GSI–$\tau_{gate}$ & $(\tau_{gate}, \tau_{sub}) = (0, 9)$" scheme, resulting in superior performance compared to "GSI–$\tau_{gate}$" again. These results demonstrate that the present FeFET RC system with substrate input possesses flexibility in exploring task-specific best combinations of input conditions. By incorporating favorable input conditions, the system can be optimized for specific tasks to achieve superior performance.

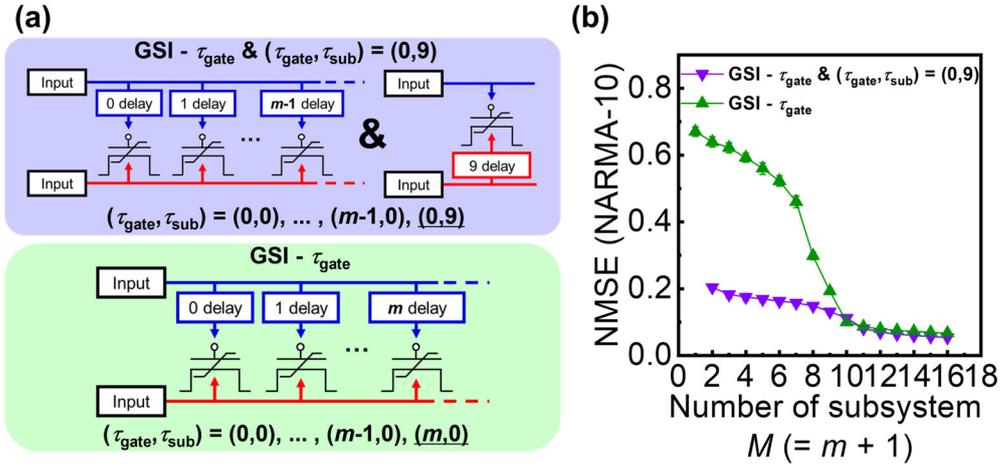

Fig.13 (a) Comparison between "GSI–$\tau_{gate}$" and "GSI–$\tau_{gate}$ & $(\tau_{gate}, \tau_{sub}) = (0,9)$". (b) Comparison of NMSE in the NARMA-10 prediction task between the two methods when the number of input conditions is equal.

As shown in the benchmarking results in Fig. 14, the present FeFET RC system achieves performance comparable to that of other physical reservoir computing systems, reaching an NMSE of 0.054 with 640 virtual nodes. While some physical RC platforms, such as photonic reservoir computing, can also achieve high performance, FeFET-based RC has a significant advantage in terms of CMOS compatibility and ease of on-chip integration. In particular, FeFETs can be integrated on a silicon chip together with peripheral circuitry, enabling compact and scalable system implementation. An array of many FeFETs can be integrate on-chip in advance, with delay conditions at each FeFET adjustable according to target applications through external circuitry. In other words, the FeFET matrix array structure proposed in this study, leveraging its high integration, can be regarded as a general-purpose reservoir computing platform.

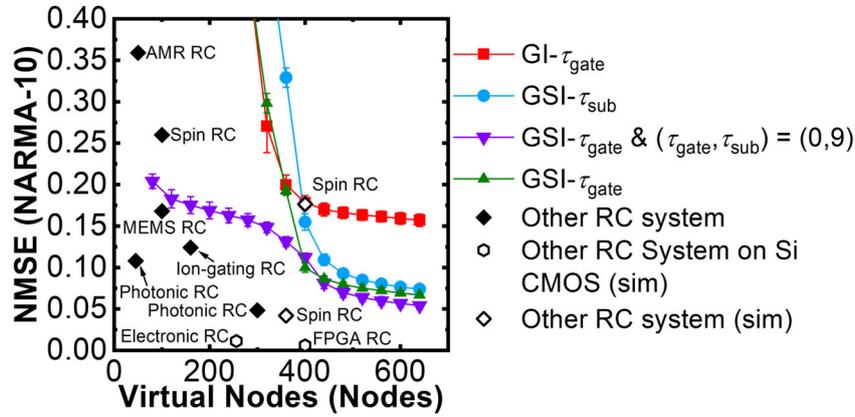

Fig.14 Benchmarking against other physical reservoir computing systems. [22–29]

**Conclusion**

We have demonstrated that the ferroelectric polarization dynamics can be controlled by applying a delayed input to the substrate. In particular, negative substrate voltage can partially suppress the polarization switching in the negative direction by controlling the depletion region and in turn the electric field across the ferroelectric layer. By incorporating this property into the FeFET RC system, better temporal nonlinearity can be introduced to the FeFET and the internal states of FeFETs can be separately controlled by both the gate and substrate terminals. Furthermore, it has been shown that, by independently controlling the timing of the gate and substrate inputs and combining FeFETs configured with various delay conditions, the FeFET RC system can perform well even with comparatively difficult tasks, including NARMA with high orders. Engineering the RC system with combinations of effective delay conditions suggests the potential to optimize the system for specific tasks with minimal system costs. This is beneficial for RC because RC is expected to be a computational framework that is particularly computationally efficient for specific tasks, rather than a universal system to deal with any tasks. These results demonstrate that introducing substrate input to FeFET, combined with the parallelization enabled by high CMOS compatibility of FeFET, enables a flexible FeFET RC system to be implemented on a single Si chip.